\documentclass[english,pra,twocolumn,groupedaddress,reprint]{revtex4}
\usepackage{graphicx}
\usepackage{amsfonts}
\usepackage[T1]{fontenc}
\usepackage[latin9]{inputenc}
\usepackage{color}
\usepackage{amsmath}
\usepackage{amssymb}
\usepackage{esint}
\usepackage{comment}
\usepackage{bbold}
\usepackage{tikz}
\usepackage{minitoc}
\usepackage{bm}
\usepackage{CJK}

\begin{document}

\title{Weak amplification and single-photon weak-coupling optomechanics}
\author{Tao Wang\footnote{suiyueqiaoqiao@163.com} $^{1,2}$ and Gang Li\footnote{ligang0311@sina.cn} $^{3}$}
\affiliation{$^{1}$College of Physics, Jilin University, Changchun 130012, People's Republic of China}
\affiliation{$^{2}$College of Physics, Tonghua Normal University, Tonghua 134000, People's Republic of China}
\affiliation{$^{3}$School of Physics and Optoelectronic Technology,
Dalian University of Technology, Dalian 116024, People's Republic of China}
\date{\today }

\begin{abstract}
The meaning and its possible applications of post-selected weak amplification in optomechanical system is concisely reviewed.~~\newline
~~~~\newline
PACS numbers: 42.50.Wk, 42.65.Hw, 03.65.Ta
\end{abstract}

\maketitle

\section{Introduction}

Two years ago (2013) T. Wang recognized that a single-photon weak-coupling optomechanics is possible if we adopted a post-selected
weak measurement scheme \cite{Aharonov88,Dresse14} and then T. Wang and G. Li found an excellent work in this field had been published in 2012 proposed by B.  Pepper, R. Ghobadi, E. Jeffrey, C. Simon and D. Bouwmeester  \cite{Bouwmeester12}. However the purpose of their work is to achieve macroscopic quantum superpositions and to generate one-phonon state in optomechanical systems and is not to amplify the tiny effect in the process.

In their paper they said: ``This aspect of our scheme is related to the weak measurement
formalism, with the optomechanical device
essentially acting as a `pointer' which weakly
measures photon number. However, it operates outside
the weak measurement regime due to its totally
orthogonal postselection.'' This means optomechanical systems provide a unique platform to
discuss weak amplification in an exact way when the post-selected state of the photon is
orthogonal to the initial state, which cannot be explained by the usual weak measurement results.

This inspired us that this discussions are exact so not only the amplification effect but also the amplification limit can be both available. This is very important for deepening our understanding of the weak amplification mechanism. Moreover this study will be useful for finding new phenomena in optomechanics in the one-photon weak-coupling regime which is often considered impossible.

Subsequently a series of works are performed by G. Li,
his collaborators and T. Wang \cite{Li1,Li2,Li3,Li4,Li5,Li6}. We discuss the Kerr phase effect in the optomechanical interaction, the phase shifter effect ,
the coherent effect, the squeezing effect and especially the thermal noise effect. However in these works some deeper meanings and their
possible applications are not fully elaborated and someone who is not familiar with
the ideas of weak amplification and single-photon weak-coupling optomechanics may misunderstand this. In this short review we illustrates the value of
our works and clarify the key components hidden in the complicated calculation.

In the first published paper  \cite{Li1}  we find that revealing the weaker effect using post-selected weak measurement is possible such as the tiny Kerr phase effect in the optomechanical interaction. If there is no post-selection the Kerr phase can be ignored. This inspires us to reveal other weaker effect using post-selected weak measurement.

However in this paper \cite{Li1} the pointer is prepared in the ground state which is not easy to prepare completely and the maximum displacement can only reach the vacuum fluctuation. For reaching larger amplification degree ultimately we let the pointer in a thermal state  \cite{Li5} which is easy to prepare and the amplification degree can reach the thermal fluctuation. Moreover G. Li found the thermal state pointer can also enormously improve the precision \cite{Li6}.

We believe the series of our works \cite{Li1,Li3,Li5,Li6} , especially using the thermal state as the pointer \cite{Li5,Li6} , provide enough toolbox to reveal weaker effect in optomechanical systems such as detecting the  gravitational wave and revealing other feeble gravitational effect.

\section{Single-photon weak-coupling optomechanics }

Light or photons can exert a force to a mirror and the displacement of the mirror is proportional to
the intensity of the light or the number of the photons in the lowest approximation (in the weak-coupling regime this approximation is enough) \cite{Marquardt13}. And the Hamiltonian
of optomechanical system is expressed as:

\begin{equation}
H=\hbar \omega _{0} a^{\dagger }a+\hbar \omega _{m}c^{\dagger
}c-\hbar ga^{\dagger }a(c^{\dagger }+c),  \label{a}
\end{equation}
where $\hbar $ is Plank's constant. $\omega _{0}$ and $a$  are
frequency and annihilation operator of the optical cavity A  ,
respectively. $c$ is annihilation operator of mechanical system (mirror) with angular
frequency $\omega _{m}$ and the optomechanical coupling strength $g=\frac{%
\omega _{0}}{L} \sigma  $, where $L$ is the length of the cavity A  . $%
\sigma =(\hbar /2m\omega _{m})^{1/2}$ is the zero-point fluctuation and $m$
is the mass of mechanical system. The maximum displacement of the mirror can reveal the photon number in the cavity A, so it is a  native measurement pointer.

If only one photon is in the cavity A the induced maximum displacement of the mirror is $4\kappa \sigma$ where  $\kappa=\frac{g}{\omega_{0}}$ \cite{Marshall03}, so if $\kappa< \frac{1}{4}$ the displacement  can not reach the vacuum fluctuation and this single-photon interaction is hard to detect.
We can have three ways to improve this. One way is using a strong light driving which can enhance the effect interaction strength between the laser and the mirror. The second way is to increase $g$ to reach the one-photon strong-coupling regime. And the third way is to adopt the idea of post-selected weak measurement \cite{Aharonov88,Dresse14}. This idea has been discussed in Pepper \emph{et al}' paper using a  March-Zehnder
interferometer in Figure 1 \cite{Bouwmeester12}. Here it is a weak measurement model where
the mirror is used as the pointer to measure the number of photon in cavity
A. In this outstanding paper their purpose is to create macroscopic quantum superpositions.
In fact this work is an optomechanical realization of the Fock-state view of weak-value measurements proposed by C. Simon and E. S. Polzik \cite{Simon11}.
Our works are excited by this idea.

\begin{figure}[b]
\includegraphics[scale=0.43]{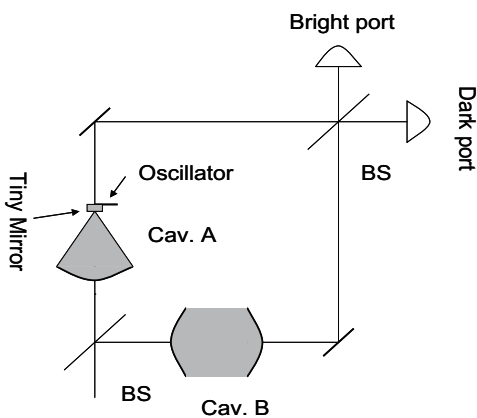}.
\caption{The photon enters the first beam splitter of March-Zehnder
interferometer with an optomechanical cavity A and a conventional cavity B.
The photon weakly affectsv the small mirror. After the second beam splitter,
dark port is detected, i.e., postselection acts on the case where the mirror
has been affected by a photon, and fails otherwise.}
\end{figure}

\section{Post-selected weak measurement}

The idea of post-selected weak measurement suggested by Y. Aharonov, D.Z. Albert, and L. Vaidman \cite{Aharonov88} is very easy. The quantum system is initially prepared in a superposition state $|\psi \rangle$ and the measurement pointer is prepared in a pure Gaussian state. The interaction between the system and the pointer is weak via a Hamiltonian $H=\chi  \sigma_{z}P$, where $\sigma_{z}$ is a spin-like observable
for the system and $P$ is a momentum-like observable for the pointer. The displacement of the pointer can reveal the spin-like observable if the interaction is strong.

For the weakness of the interaction the total system can not change too much and generally speaking this is meaningless. However if we add a post-selection of the state $ |\phi \rangle$ of the quantum system which is nearly orthogonal to the initial state $|\psi \rangle$ there can exist a distinguished state and this can offer an amplification effect.

In \cite{Simon11} they said: ``Any Gaussian can be seen as the ground state of a fictional harmonic oscillator Hamiltonian.'' and they can verify that the amplification effect results from the equal superposition of the ground state $|0 \rangle$ and the first excited states $|1 \rangle$. This state needs the final state of the quantum system  nearly orthogonal to the initial state. If the final state is absolutely orthogonal to the initial state the pointer will be in the first excited states $|1 \rangle$. This is just the generated macroscopic quantum superpositions in \cite{Bouwmeester12}. So this inspired T. Wang to discuss the amplification effect in a real harmonic oscillator system and we think optomechanical system is a good choice for this aim and then we found this work had been elaborated in \cite{Bouwmeester12}. 

If we discuss a real quantum pointer the Fock-state view is an immediate calculation approach so this connection is often ignored. The Fock-state view bridges between the old way discussing the weak measurement and the new applications with a real quantum system as the measurement pointer.

So whether a quantum pointer gains an advantage over the classical one is an interesting question which is the primary motivation of our work. However this is abandoned by us subsequently.

\section{Amplifying tiny Kerr phase effect and the amplification limit}

The optomechanical interaction (1) can induce two effects: one is the periodic displacement of the mirror and another is the Kerr phase shift \cite{Li1}. The Kerr phase shift is proportional to the square of $\kappa$. If the $\kappa$ is much smaller than unitary the Kerr phase shift can be ignored. This is the way done in \cite{Bouwmeester12}. What we are curious is whether the tiny Kerr phase can manifest itself in the post-selected weak measurement scheme.

The results are very interesting. The first one is that through exact calculation the maximum displacement can reach the vacuum fluctuation $\sigma$ which is not related to $\kappa$. The second one is when the final state of the photon is orthogonal to the initial state the Kerr phase effect can occur and the maximum displacement can also reach the vacuum fluctuations $\sigma$. Both of the two results means any tiny effect may be amplified to the vacuum fluctuations $\sigma$. This results are very important to the idea of post-selected weak amplification.

In the previous studies we have known the weak amplification effect but we don't know the extent it can be. If amplification effect is small it is of no use. Determining the amplification limit is very important for this scheme.

This question is also generally studied by S. S. Pang, T. A. Brun, S. J. Wu and Z. B. Chen \cite{Pang14} which is published in the same issue of the journal PRA with our first paper \cite{Li1}. They discussed the amplification limit of weak measurements with a variational approach. They derived a
formal asymptotic solution for a weak-coupling Hamiltonian and revealed the surprising
property that the solution is independent of the coupling strength depending only on the initial state of
the detector and the dimension of the system or the detector. They said: ``If one wants to enlarge the
maximal position shift of the detector, one should choose an initial state with wider spread for the detector; to enlarge the
maximal momentum shift of the detector, the initial spread of the detector should be narrower. The width of the initial
detector wave function decides the amplification limit of a weak measurement with that detector.'' Although they also discussed the mixed state, the results  based on the pure state.

We are sorry that this paper only attracts one of us (T. Wang) recently. The aim of our works is applying weak amplification in the optomechanical system to reveal the weaker effect in the process. Because the calculation is exact so the amplification limit can also be given. And importantly in our model the maximum displacement induced by the weaker effect is also the vacuum fluctuations $\sigma$. This is not discussed in \cite{Pang14}.

These works indicate an incredible result. The amplification limit for postselected weak measurement is robust to the wavefunctions of the pointer and the weak interaction. 

\section{Coherent state and Squeezing state as the pointer}

Fock-state view inspired two kinds of works. One is to find whether the non-classical states are superior to the pure Gaussian state used in the old way and another is whether we can improve the amplification limit.

For the first kind, we discuss two situations where the pointer is prepared in a coherent state \cite{Li2} and in a squeezed state \cite{Li3}. We find the coherent state as a pointer can also induce amplification but the maximum displacement is also the vacuum fluctuation $\sigma$ and if we use coherent state the amplification can also occur when the final state of the photon is orthogonal to the initial state which is due to the noncommutativity of quantum mechanics and can not be explained by the classical one.

However we find the squeezed state and the coherent squeezed state can improve the amplification limit and the maximum displacement of the mirror can reach $e^{r} \sigma$ where $r$ is the squeezing parameters. In fact $e^{r} \sigma$ is the fluctuations of the quadrature component with increased fluctuations in the
squeezed state. This is a rare situation  taking advantage of the enhanced quadrature. This result is very interesting but is difficult to realize in the optomechanical system.

\section{Thermal state as the pointer}

After these works we realized that in 2014 the maximum displacement of the pointer in the post-selected weak measurement may be decided by the fluctuation of the pointer. This meaning is also shown in \cite{Pang14}. This inspired us to let the pointer in the thermal state \cite{Li5} and we find that the maximal value of this effect can reach the thermal fluctuations  $  (\frac{1+z}{1-z})^{1/2}\sigma $, where $z=e^{-\hbar\omega _{m}/k_{B}T}$, $k_{B}$ is the Boltzmann constant and $T$ is the temperature. This is a very counter-intuitional and powerful result.

If there is no postselection the one-photon induced displacement is $4\kappa \sigma$ which is not related to the style of the pointer. However in the weak amplification scheme the style of the pointer is important due to the interference effect. It is the width or fluctuations of the initial state of the pointer to determine the maximum amplification degree.

This result seems very easy but it is beyond our imagination. In fact there is no clue for this result and only an exact result can give it. Thermal state is the most classical state and can be diagonal in the Fock state basis. Our calculations reals that weak amplification can be realized for any number state. 

For optomechanical system thermal state is very easy to prepare and  pre-cooling the mirror to the ground state is not necessary. This enormously reduces the conditions to apply the weak amplification in the optomechanical system. And in room temperature the maximum amplification degree can be nearly one hundred  thousands times than the vacuum fluctuations and this tremendously reduces the detection difficulty.

Thus we provide enough toolbox to reveal the weaker effect in optomechanical system and we realize that thermal state is the best choice for a real oscillator for weak amplification.

\section{Precision}

Recently whether weak amplification can improve the measurement precision attracts much debates \cite{Knee14}. However their arguments bases on the pure Gaussian state. S. S. Pang and T. A. Brun found that the presence of ``nonclassicality'' in the pointer states can improve the precision \cite{Pang15} and G. Li found that using thermal state as the pointer can enormously improve the precision \cite{Li6}.

So we think weak amplification scheme is very different from the previous amplification schemes. And the result of G. Li lay a solid foundation for weak amplification in optomechanical system.

\section{Amplifying faint gravitational effect }

At the beginning the ultimate goal of our works is to design new scheme to detect the faint gravitational effect in the optomechanical setup using the idea of post-selected weak measurement such as gravitational wave. We think the two papers \cite{Li5,Li6} provide enough toolbox for this ultimate goal. The scheme will be opposite from the previous ones which need not exclude all the disturbance and noise for the gravitational effect. We think this can greatly reduce the difficulty of gravitational wave. This may be the most important application of our works.

\section{Conclusion}

In the short review of our works we try to clarify some deeper meanings of our work and its possible application in detecting gravitational effect. However we only highlight the main results of our work and many details and related works are not mentioned. We expect a detailed longer review will be given in future.

\end{document}